\documentclass[a4paper]{jpconf}

\usepackage{graphicx}
\usepackage{amsfonts,amsmath,amssymb,amsthm,latexsym}
\usepackage[all]{xy}
\usepackage{hyperref}

\begin{document}

\title{Graded Fock--like representations for a system of algebraically interacting paraparticles}

\author{Konstantinos Kanakoglou$^{1,2}$, Alfredo Herrera--Aguilar$^1$}

\address{$^1$ Instituto de F\'{\i}sica y Matem\'{a}ticas (\textsc{Ifm}),
Universidad Michoacana de San Nicol\'{a}s de Hidalgo (\textsc{Umsnh}),
Edificio C-3, Cd. Universitaria, CP 58040, Morelia, Michoac\'{a}n, \textsc{Mexico} \\
$^2$ School of Physics,
Nuclear and Elementary Particle Physics Department, Aristotle {\scriptsize University}
of Thessaloniki (\textsc{Auth}),
54124, Thessaloniki, \textsc{Greece}}

\ead{kanakoglou@ifm.umich.mx, herrera@ifm.umich.mx}

\begin{abstract}
We will present and study an algebra describing a mixed paraparticle
model, known in the bibliography as ``The Relative Parabose Set
(\textsc{Rpbs})''. Focusing in the special case of a single
parabosonic and a single parafermionic degree of freedom
$P_{BF}^{(1,1)}$, we will construct a class of Fock--like
representations of this algebra, dependent on a positive parameter
$p$ a kind of \emph{generalized parastatistics order}. Mathematical
properties of the Fock--like modules will be investigated for all
values of $p$ and constructions such as ladder operators,
irreducibility (for the carrier spaces) and $(\mathbb{Z}_{2} \times
\mathbb{Z}_{2})$--gradings (for both the carrier spaces and the
algebra itself) will be established.
\end{abstract}

\section{Introduction}  \label{sect1}

Our central object of study in this short letter, will be the
\emph{Relative Parabose Set algebra} $P_{BF}^{(1,1)}$ in a single
parabosonic and a single parafermionic degree of freedom. It belongs
to the general family of paraparticle algebras intimately related to
the Wigner quantization scheme \cite{Wi}. The ``\textit{free}"
paraparticle algebras (parabosonic and parafermionic algebras) have
been -implicitly- introduced in the early '50's \cite{Green} while
the ``\textit{mixed}" paraparticle models such as $P_{BF}$ (and a
couple of others as well) have been first introduced in
\cite{GreeMe}.

The algebra $P_{BF}^{(1,1)}$ is generated (as an algebra over $\mathbb{C}$) by the four generators $b^{+}, b^{-}, f^{+}, f^{-}$ subject to the
usual trilinear relations of the \textit{free parabosonic} and the \textit{free parafermionic} algebras which can be compactly summarized as
\begin{equation} \label{equat1}
\begin{array}{ccc}
\big[ \{ b^{\xi},  b^{\eta}\}, b^{\epsilon}  \big] = (\epsilon - \eta)b^{\xi} + (\epsilon - \xi)b^{\eta}    & , &
\big[ [ f^{\xi},  f^{\eta} ], f^{\epsilon}  \big] = \frac{1}{2}(\epsilon - \eta)^{2} f^{\xi} - \frac{1}{2}(\epsilon - \xi)^{2} f^{\eta}
 \end{array}
\end{equation}
for all values $\xi, \eta, \epsilon = \pm$, \ together with the \textit{mixed trilinear relations}
\begin{equation} \label{equat2}
\begin{array}{ccc}
\big[ \{ b^{\xi},  b^{\eta}\}, f^{\epsilon}  \big] = \big[ [ f^{\xi},  f^{\eta} ], b^{\epsilon}  \big] = 0,  & \big[ \{ f^{\xi},  b^{\eta}\}, b^{\epsilon}  \big] = (\epsilon - \eta) f^{\xi},  & \big\{ \{ b^{\xi},  f^{\eta}\}, f^{\epsilon}  \big\} = \frac{1}{2}(\epsilon - \eta)^{2} b^{\xi}
\end{array}
\end{equation}
for all values $\xi, \eta, \epsilon = \pm$, \ which represent a kind of algebraically established interaction between parabosonic and parafermionic degrees of freedom and \textit{characterize the relative parabose set}.

It is easy for one to see that when all combinations of the $\xi,
\eta, \epsilon = \pm$ are taken into account, the first of equations
\eqref{equat1} produces $6$ relations, the second of eq.
\eqref{equat1} produces $2$ relations and equations \eqref{equat2}
produce $24$ relations.
One can easily see (see also the discussion in \cite{KaHa3}) that
not all of these $24+6+2=32$ are algebraically independent. However,
we keep the relations in the form given in equations \eqref{equat1},
\eqref{equat2} and we do not proceed in further simplifying them,
because of the compact notational and computational advantages of
this presentation. (See also \cite{KaDaHa1, KaHa3, Ya1}).

The purpose of this letter, will be to present the construction of a
class of representations for the relations \eqref{equat1},
\eqref{equat2}. We will present an infinite class of irreducible,
$(\mathbb{Z}_{2} \times \mathbb{Z}_{2})$--graded representations of
these relations. This class will be parametrized by the values of a
positive integer $p$ (similarly to the way that the free parabosonic
and parafermionic algebra Fock--like spaces are parametrized by the
values of a positive integer). The parameter $p$ will be called
\textit{generalized parastatistics order}. It will be shown that the
$b^{\pm}, f^{\pm}$ generators act as ``ladder" operators
(\textit{Creation/Annihilation} operators) in a kind of
two--dimensional ``ladder", generalizing thus the way that the usual
Canonical Commutation--Anticommutation relations (CCR--CAR) are
represented in their Fock spaces and also the way that the single
degree of freedom free parabosonic and free parafermionic algebras
act in their own Fock--spaces \cite{LiStVdJ} respectively.

\section{Structure of the Fock--like spaces}  \label{sect2}

In this section we are going to briefly review the results of
\cite{Ya2} on the construction of the Fock--like spaces, which will
serve as carrier spaces for the Fock--like representations of
$P_{BF}^{(1,1)}$.

An old conjecture \cite{GreeMe} on the study of the representations
of the paraparticle algebras, states that if we consider
representations of $P_{BF}^{(1,1)}$, satisfying the adjointness
conditions $(b^{-})^{\dagger} = b^{+}$ and $(f^{-})^{\dagger} =
f^{+}$, on a complex, infinite dimensional, pre--Hilbert
\footnote{in the sense that it is an inner product space, but not
necessarily complete with respect to the inner product.} space,
possessing a unique vacuum vector $|0 \rangle$ satisfying $b^{-} |0
\rangle = f^{-} |0 \rangle = b^{-} f^{+} |0 \rangle = f^{-} b^{+} |0
\rangle = 0$ and $\langle0|0\rangle=1$ then the following additional
conditions (where $p$ may be an arbitrary positive integer)
\begin{equation}  \label{singleoutFock}
b^{-} b^{+} |0 \rangle = f^{-} f^{+} |0 \rangle = p |0 \rangle
\end{equation}
single out an irreducible representation which is unique up to unitary equivalence.
In other words the above statement, tells us that for any positive integer $p$ there is an irreducible representation of $P_{BF}^{(1,1)}$ uniquely specified (up to unitary equivalence) by the above relations.

We now proceed in summarizing the results of \cite{Ya2}: The carrier
spaces of the Fock--like representations of $P_{BF}^{(1,1)}$ are
$\bigoplus_{n=0}^{p} \bigoplus_{m=0}^{\infty} \mathcal{V}_{m,n}$
where $p$ is an arbitrary (but fixed) positive integer. The
subspaces $\mathcal{V}_{m,n}$ are 2--dim except for the cases $m =
0$, $n = 0, p$ i.e. except the subspaces $\mathcal{V}_{0,n}$,
$\mathcal{V}_{m,0}$, $\mathcal{V}_{m,p}$ which are 1--dim. Let us
see how the corresponding vectors look like:

$\bullet$ \underline{If $0 < m$ and $0 < n < p$}, then the subspace
$\mathcal{V}_{m,n}$ is spanned by all vectors of the form {\small
\begin{equation}   \label{spanvect}
\Big| \begin{array}{c}
      m_{1}, m_{2}, ..., m_{l}  \\
n_{0},n_{1}, n_{2}, ..., n_{l}
      \end{array} \Big\rangle \equiv (f^{+})^{n_{0}} (b^{+})^{m_{1}} (f^{+})^{n_{1}} (b^{+})^{m_{2}} (f^{+})^{n_{2}} ... (b^{+})^{m_{l}} (f^{+})^{n_{l}} |0 \rangle
\end{equation}
}
where $m_{1} + m_{2} + ... + m_{l} = m$ , $n_{0} + n_{1} + n_{2} + ... + n_{l} = n$ and $m_{i} \geq 1$ (for $i = 1, 2, ... , l$), $n_{i} \geq 1$ (for $i = 1, 2, ..., l-1$) and $n_{0}, n_{l} \geq 0$.

For any specific combination of values $(m,n)$ the corresponding subspace
$\mathcal{V}_{m,n}$ has a basis consisting of the two vectors
($R^{\eta} = \frac{1}{2} \{ b^{\eta}, f^{\eta} \}$ for $\eta = \pm$)
{\footnotesize
\begin{equation} \label{subspbase}
\begin{array}{cc}
|m,n,\alpha\rangle\equiv(f^{+})^{n}(b^{+})^{m}|0\rangle=
\Big|\!\begin{array}{c}
m  \\
n,0
\end{array}\!\Big\rangle,  &
|m,n,\beta\rangle\equiv(f^{+})^{(n-1)}(b^{+})^{(m-1)}
R^{+}|0\rangle=\frac{1}{2}\Big|\!\begin{array}{c}
m  \\
n-1,1
\end{array}\!\Big\rangle+
\frac{1}{2}\Big|\!\begin{array}{c}
                  m-1,1  \\
                  n-1,1,0
                  \end{array}\!\Big\rangle
\end{array}
\end{equation}
}

$\bullet$ \underline{If $m = 0$ or $n = 0, p \ $}, the vectors $|0, n, \beta \rangle$ and $|m, 0, \beta \rangle$ are (by definition) zero and furthermore, for $m\neq0$, the vector $|m, p, \beta \rangle$ becomes parallel to $|m, p, \alpha \rangle$: $|m, p, \beta \rangle=\frac{1}{p} |m, p, \alpha \rangle$

$\bullet$ \underline{If $n \geq p+1$}, all basis vectors of \eqref{subspbase} vanish.

\emph{Note 1:} The above described subspaces of $\bigoplus_{n=0}^{p} \bigoplus_{m=0}^{\infty} \mathcal{V}_{m,n}$ can be visualized as follows:
\qquad \qquad \qquad \qquad
{\small
$
\begin{array}{cccccccc}
    \mathcal{V}_{0,0} & \mathcal{V}_{0,1} & \ldots  &  \mathcal{V}_{0,n} & \ldots & \ldots & \mathcal{V}_{0,p-1} & \mathcal{V}_{0,p}  \\
    \mathcal{V}_{1,0} & \mathcal{V}_{1,1} & \ldots & \mathcal{V}_{1,n} & \ldots &  \ldots & \mathcal{V}_{1,p-1} & \mathcal{V}_{1,p}  \\
    \vdots & \vdots & \ldots  & \vdots & \ldots & \ldots  & \vdots & \vdots     \\
    \mathcal{V}_{m,0} & \mathcal{V}_{m,1} & \ldots & \mathcal{V}_{m,n} & \mathcal{V}_{m,n+1} & \ldots & \ldots & \mathcal{V}_{m,p} \\
    \vdots & \vdots &  \ldots & \mathcal{V}_{m+1,n} & \ldots &  \ldots  & \vdots  & \vdots     \\
    \vdots & \vdots & \ldots & \vdots & \ldots & \ldots  & \vdots & \vdots
\end{array}
$
}

$\bullet$ Inside each one of the above presented $2$--dim subspaces $\mathcal{V}_{m,n}$
(with $m\neq0$, $n\neq0,p$) the vectors of \eqref{subspbase} are linearly independent
and constitute a basis. However, these vectors are neither orthogonal nor normalized.
We can show that an orthonormal set of basis vectors can be obtained as \\
\qquad \qquad \qquad
$
\begin{array}{cccc}
| m, n, + \rangle = c_{+} | m, n, \alpha \rangle &  &  & | m, n, - \rangle = -c_{-} \Big( | m, n, \alpha \rangle - p | m, n, \beta \rangle \Big)
\end{array}
$ \\
where $c_{\pm}$ are suitable normalization factors \cite{Ya2}. Now we can show orthonormalization $\langle m,n,s|m^{'},n^{'},s^{'}\rangle=\delta_{m,m^{'}}\delta_{n,n^{'}}\delta_{s,s^{'}}$ and completeness $\sum_{m=0}^{\infty} \sum_{n=0}^{p} \sum_{s=\pm}|m,n,s\rangle \langle m,n,s| = 1$.

\emph{Note 2:} If we consider the Hermitian operators $N_{s} = \frac{1}{p} \big( N_{f}^{2} -(p+1)N_{f} + f^{+}f^{-} + \frac{p}{2} \big)$, $N_{f}=\frac{1}{2}[f^{+},f^{-}]+\frac{p}{2}$ and $N_{b} = \frac{1}{2}\{b^{+},b^{-} \}-\frac{p}{2}$ we can show that they constitute a \textit{Complete Set of Commuting Observables} (\textit{C.S.C.O.}): We have $[N_{b}, N_{f}]=[N_{b}, N_{s}]=[N_{f}, N_{s}]=0$; their common eigenvectors are exactly the elements of the orthonormal basis formerly described. Any vector $|m,n,s\rangle$ is uniquely determined as an eigenvector of $N_{b}$, $N_{f}$, $N_{s}$ by its eigenvalues $0 \leq m$, $0 \leq n \leq p$ and $s=\pm\frac{1}{2}$ respectively.

\section{Main results: Construction of the Fock--like representations} \label{sect3}

For detailed computations and proofs of the results presented in this section see \cite{KaHa3}.

\subsection{Construction of ladder operators}

We now present the formulae describing explicitly the action of the
generators (and hence of the whole algebra) of $P_{BF}^{(1,1)}$ on
the carrier spaces $\bigoplus_{n=0}^{p} \bigoplus_{m=0}^{\infty}
\mathcal{V}_{m,n}$ for any positive integer $p$: {\small
\begin{equation}  \label{equat3}
\begin{array}{l}
\blacktriangledown b^{-} \cdot | m, n, \alpha \rangle = \left\{%
\begin{array}{ll}
(-1)^{n}m | m-1, n, \alpha \rangle - 2(-1)^{n}nm | m-1, n, \beta \rangle, \ \underline{m:even}   \\   \\
-(-1)^{n} \big(2n-m-(p-1) \big) | m-1, n, \alpha \rangle - 2(-1)^{n}n(m-1) | m-1, n, \beta \rangle, \  \underline{m:odd}
\end{array}
\right.  \\
\blacktriangledown b^{-} \cdot | m, n, \beta \rangle = \left\{%
\begin{array}{ll}
-(-1)^{n} | m-1, n, \alpha \rangle + (-1)^{n}\big( 2n-m-p \big) | m-1, n, \beta \rangle, \  \underline{m:even}   \\   \\
-(-1)^{n} | m-1, n, \alpha \rangle - (-1)^{n}(m-1) | m-1, n, \beta \rangle, \  \underline{m:odd}
\end{array}
\right.    \\
  \\
\blacktriangledown f^{-} \cdot | m, n, \alpha \rangle = n(p+1-n) | m, n-1, \alpha \rangle, \blacktriangledown b^{+} \cdot | m, n, \alpha \rangle = (-1)^{n} | m+1, n, \alpha \rangle - (-1)^{n}2n | m+1, n, \beta \rangle  \\
\blacktriangledown f^{-} \cdot | m, n, \beta \rangle =  | m, n-1, \alpha \rangle + (n-1)(p-n)| m, n-1, \beta \rangle,  \quad \blacktriangledown b^{+} \cdot | m, n, \beta \rangle = -(-1)^{n} | m+1, n, \beta \rangle \\  \\
\!\!
\begin{array}{lll}
\blacktriangledown f^{+} \! \cdot \! | m, n, \alpha \rangle = \left\{%
\begin{array}{cc}
| m, n+1, \alpha \rangle,  \underline{if \ n \leq p-1}   \\
0, \qquad \qquad \underline{if \ n \geq p}
\end{array}
\right.
   &    &
\blacktriangledown f^{+} \! \cdot \! | m, n, \beta \rangle = \left\{%
\begin{array}{cc}
| m, n+1, \beta \rangle,  \underline{if \ n \leq p-1}   \\
0, \qquad \qquad \underline{if \ n \geq p}
\end{array}
\right.
\end{array}
\end{array}
\end{equation}
} for all integers $0 \leq m$, $0 \leq n \leq p$. The direct proof
\cite{KaHa3} of these formulae involves lengthy ``normal--ordering"
algebraic computations inside $P_{BF}^{(1,1)}$. We must take into
account: {\small $(a).$} the relations \eqref{equat1},
\eqref{equat2} of $P_{BF}^{(1,1)}$, {\small $(b).$} the relations
\eqref{singleoutFock} together with {\small
$b^{-}|0\rangle\!=\!f^{-}|0\rangle\!=\!b^{-}
f^{+}|0\rangle\!=\!f^{-}b^{+}|0\rangle\!=\!0$} and {\small $(c).$}
the structure and properties of the corresponding carrier space as
described in Sect. \ref{sect2}.

Apart from the direct proof, eq. \eqref{equat3} have also been
checked and verified via the
\href{http://homepage.cem.itesm.mx/lgomez/quantum/}{Quantum}
\cite{GMFD} add--on for Mathematica 7.0, which is an add-on for
performing symbolic algebraic computations, including the use of
generalized Dirac notation. What we have actually verified via the
use of this package, is that the actions formulae \eqref{equat3}
preserve all of the relations \eqref{equat1}, \eqref{equat2} of
$P_{BF}^{(1,1)}$.

\subsection{Irreducibility}

In the following diagram we provide a ``visual" interpretation of
relations \eqref{equat3} i.e. of the action of the generators of
$P_{BF}^{(1,1)}$ on the direct summand subspaces of the Fock-like
carrier space $\bigoplus_{n=0}^{p} \bigoplus_{m=0}^{\infty}
\mathcal{V}_{m,n}$. We can easily figure out that we have a kind of
\textit{generalized creation--annihilation operators} acting on a
two dimensional ladder of subspaces: {\scriptsize
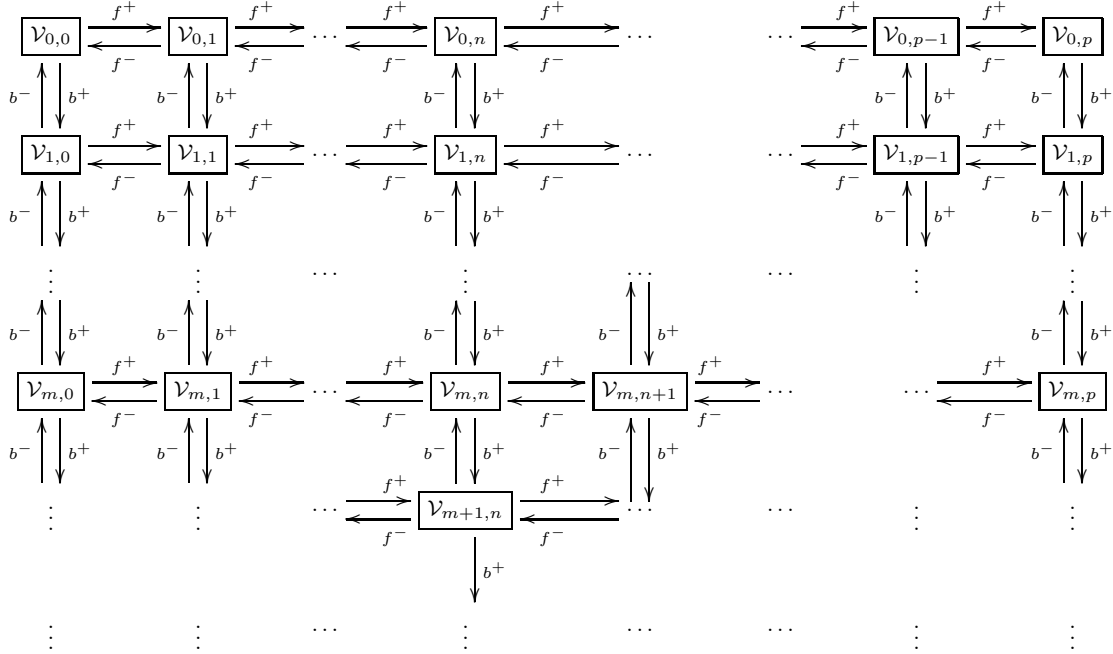
\xymatrix{
    \boxed{\mathcal{V}_{0,0}} \ar@<1ex>[r]^{f^{+}} \ar@<1ex>[d]^{b^{+}} & \boxed{\mathcal{V}_{0,1}} \ar@<1ex>[r]^{f^{+}} \ar@<1ex>[l]^{f^{-}} \ar@<1ex>[d]^{b^{+}} & \ldots \ar@<1ex>[r]^{f^{+}} \ar@<1ex>[l]^{f^{-}} &  \boxed{\mathcal{V}_{0,n}} \ar@<1ex>[r]^{f^{+}} \ar@<1ex>[l]^{f^{-}} \ar@<1ex>[d]^{b^{+}} & \ldots \ar@<1ex>[l]^{f^{-}} & \ldots \ar@<1ex>[r]^{f^{+}} & \boxed{\mathcal{V}_{0,p-1}} \ar@<1ex>[r]^{f^{+}} \ar@<1ex>[l]^{f^{-}} \ar@<1ex>[d]^{b^{+}} & \boxed{\mathcal{V}_{0,p}} \ar@<1ex>[l]^{f^{-}} \ar@<1ex>[d]^{b^{+}} \\
    \boxed{\mathcal{V}_{1,0}} \ar@<1ex>[r]^{f^{+}} \ar@<1ex>[d]^{b^{+}} \ar@<1ex>[u]^{b^{-}} & \boxed{\mathcal{V}_{1,1}} \ar@<1ex>[r]^{f^{+}} \ar@<1ex>[l]^{f^{-}} \ar@<1ex>[d]^{b^{+}} \ar@<1ex>[u]^{b^{-}} & \ldots \ar@<1ex>[r]^{f^{+}} \ar@<1ex>[l]^{f^{-}} &  \boxed{\mathcal{V}_{1,n}} \ar@<1ex>[r]^{f^{+}} \ar@<1ex>[l]^{f^{-}} \ar@<1ex>[d]^{b^{+}} \ar@<1ex>[u]^{b^{-}} & \ldots \ar@<1ex>[l]^{f^{-}} &  \ldots \ar@<1ex>[r]^{f^{+}} & \boxed{\mathcal{V}_{1,p-1}} \ar@<1ex>[r]^{f^{+}} \ar@<1ex>[l]^{f^{-}} \ar@<1ex>[d]^{b^{+}} \ar@<1ex>[u]^{b^{-}} & \boxed{\mathcal{V}_{1,p}} \ar@<1ex>[d]^{b^{+}} \ar@<1ex>[u]^{b^{-}} \ar@<1ex>[l]^{f^{-}} \\
    \vdots \ar@<1ex>[d]^{b^{+}} \ar@<1ex>[u]^{b^{-}} & \vdots \ar@<1ex>[d]^{b^{+}} \ar@<1ex>[u]^{b^{-}} & \ldots  & \vdots \ar@<1ex>[d]^{b^{+}} \ar@<1ex>[u]^{b^{-}}  & \ldots \ar@<1ex>[d]^{b^{+}}  & \ldots  & \vdots \ar@<1ex>[u]^{b^{-}}  & \vdots \ar@<1ex>[d]^{b^{+}} \ar@<1ex>[u]^{b^{-}}    \\
    \boxed{\mathcal{V}_{m,0}} \ar@<1ex>[r]^{f^{+}} \ar@<1ex>[d]^{b^{+}} \ar@<1ex>[u]^{b^{-}}  & \boxed{\mathcal{V}_{m,1}} \ar@<1ex>[r]^{f^{+}} \ar@<1ex>[l]^{f^{-}} \ar@<1ex>[d]^{b^{+}} \ar@<1ex>[u]^{b^{-}} & \ldots \ar@<1ex>[r]^{f^{+}} \ar@<1ex>[l]^{f^{-}} & \boxed{\mathcal{V}_{m,n}} \ar@<1ex>[r]^{f^{+}} \ar@<1ex>[l]^{f^{-}} \ar@<1ex>[d]^{b^{+}} \ar@<1ex>[u]^{b^{-}} & \boxed{\mathcal{V}_{m,n+1}} \ar@<1ex>[r]^{f^{+}} \ar@<1ex>[l]^{f^{-}} \ar@<1ex>[d]^{b^{+}} \ar@<1ex>[u]^{b^{-}} & \ldots \ar@<1ex>[l]^{f^{-}}  & \ldots \ar@<1ex>[r]^{f^{+}} & \boxed{\mathcal{V}_{m,p}} \ar@<1ex>[l]^{f^{-}} \ar@<1ex>[d]^{b^{+}} \ar@<1ex>[u]^{b^{-}} \\
    \vdots \ar@<1ex>[u]^{b^{-}} & \vdots \ar@<1ex>[u]^{b^{-}} &  \ldots \ar@<1ex>[r]^{f^{+}}  & \boxed{\mathcal{V}_{m+1,n}} \ar@<1ex>[r]^{f^{+}} \ar@<1ex>[l]^{f^{-}} \ar@<1ex>[d]^{b^{+}} \ar@<1ex>[u]^{b^{-}} & \ldots \ar@<1ex>[l]^{f^{-}} \ar@<1ex>[u]^{b^{-}} &  \ldots  & \vdots  & \vdots \ar@<1ex>[u]^{b^{-}}    \\
    \vdots & \vdots & \ldots & \vdots & \ldots & \ldots  & \vdots & \vdots
}
}  \\
Initiating from the remark that we are dealing with a cyclic module
which moreover can be generated by any of its elements, we can prove
\cite{KaHa3} that the Fock--like representations, explicitly given
by \eqref{equat3} and visually represented in the above diagram, are
irreducible representations (or: simple $P_{BF}^{(1,1)}$-modules)
for any $p \in \mathbb{N}^{*}$.

\subsection{Klein--group Grading of the representation}

Defining $\verb"deg"|m,n,\alpha\rangle\!=\!\verb"deg"|m,n,\beta
\rangle\!=\!\big(m\ \textsf{mod}\ 2,\ n\ \textsf{mod}\ 2
\big)\!\in\!\mathbb{Z}_{2}\!\times\!\mathbb{Z}_{2}$ for the carrier
spaces and $\verb"deg" b^{\pm}\!=\!(1,0)$, $\verb"deg" f^{\pm}\!=\!
(0,1)$ for the algebra, the Fock--like representations of
$P_{BF}^{(1,1)}$ over $\bigoplus_{n=0}^{p} \bigoplus_{m=0}^{\infty}
\mathcal{V}_{m,n}$, become $(\mathbb{Z}_{2} \times
\mathbb{Z}_{2})$--graded modules, $\forall p\in\mathbb{N}^{*}$. (For
proof and details see \cite{KaHa3}).

\ack

{\small KK would like to thank the whole staff of \textsc{Ifm},
\textsc{Umsnh} for providing a challenging and stimulating
atmosphere while preparing this article. His work was supported by
the research project \textsc{Conacyt}/No. J60060. AHA was supported
by \textsc{Cic} 4.16 and \textsc{Conacyt}/No. J60060; he is also
grateful to \textsc{Sni}. }

\end{document}